\documentclass[3p,twocolumn, review]{elsarticle}

\usepackage[utf8]{inputenc}
\usepackage{graphicx}
\usepackage{xcolor}
\usepackage{mathptmx}
\usepackage{amsfonts,amsmath, amsthm}
\usepackage{algorithm}
\usepackage{algpseudocode}
\usepackage{multirow}
\usepackage{comment}
\usepackage{bbm}
\usepackage{placeins}
\usepackage{hyperref}

\graphicspath{images/}

\algdef{SE}[DOWHILE]{Do}{doWhile}{\algorithmicdo}[1]{\algorithmicwhile\ #1}%

\usepackage{amssymb}

\journal{Computational Science}

\begin{document}

\begin{frontmatter}



\title{AVIDA: An Alternating method for Visualizing and Integrating Data}


\author[inst1]{Kathryn Dover}
\author[inst2]{Zixuan Cang}
\author[inst1]{Anna Ma}
\author[inst1,inst3]{Qing Nie\corref{cor1}}
\ead{qnie@uci.edu}
\author[inst1]{Roman Vershynin\corref{cor1}}
\ead{rvershyn@uci.edu}
\cortext[cor1]{Corresponding author.}

\affiliation[inst1]{organization={Department of Mathematics},
            addressline={University of California, Irvine}, 
            city={Irvine},
            postcode={92697}, 
            state={CA},
            country={USA}}

\affiliation[inst2]{organization={Department of Mathematics, Center for Research in Scientific Computation},
            addressline={North Carolina State University}, 
            city={Raleigh},
            postcode={27695}, 
            state={NC},
            country={USA}}
            
\affiliation[inst3]{organization={Department of Developmental \& Cell Biology, Center for Multiscale Cell Fate Research},
            addressline={University of California, Irvine}, 
            city={Irvine},
            postcode={92697}, 
            state={CA},
            country={USA}}

\begin{abstract}
High-dimensional multimodal data arise in many scientific fields. The integration of multimodal data becomes challenging when there is no known correspondence between the samples and the features of different datasets. To tackle this challenge, we introduce \emph{AVIDA}, a framework for simultaneously performing data alignment and dimension reduction. In the numerical experiments, Gromov-Wasserstein optimal transport and t-distributed stochastic neighbor embedding are used as the alignment and dimension reduction modules respectively. We show that by alternating dimension reduction and alignment, AVIDA aligns the representations of high-dimensional datasets without common features with four synthesized datasets and two real multimodal single-cell datasets. Compared to several existing methods, we demonstrate that AVIDA better preserves structures of individual datasets, especially distinct local structures in the joint low-dimensional representation, while achieving comparable alignment performance. Such a property is important in multimodal single-cell data analysis as some biological processes are uniquely captured by one of the datasets. In general applications, other methods can be used for the alignment and dimension reduction modules.\end{abstract}


\begin{highlights}
\item We propose a framework for the simultaneous alignment and dimension reduction of coupled high-dimensional datasets without any knowledge of common features.
\item By accomplishing both the dimension reduction and alignment simultaneously, AVIDA is better able to preserve fine-scale structures unique to individual data sets compared to previous approaches in data alignment.
\item To demonstrate the efficacy of our framework, we use the Gromov-Wasserstein optimal transport (GW-OT) for alignment and the t-stochastic neighbor embedding (t-SNE) for dimension reduction and visualization.
\end{highlights}


\end{frontmatter}


\section{Introduction}\label{sec1}

Databases are expanding not only in size but also with increasing complexity. In many applications, multiple measurements of a system are taken across different samples or in different feature spaces which produce \textit{multimodal data} such as texts attached to images \cite{lahat2015multimodal}. Multimodality allows a more comprehensive investigation of a system. Establishing connections among the modalities is the foundation of coherent analysis. Recently, the emerging multimodal single-cell omics has become a powerful tool to analyze different aspects of a biological system at the same time \cite{zhu2020single}. Fusing multimodal single-cell data is especially challenging when there is no direct correspondence between the measurements and the samples.

Single-cell RNA sequencing (scRNA-seq) is a recent technology that measures RNA abundance at transcriptomics level with single-cell resolution \cite{svensson2018exponential}. The maturation of the technology allows analysis with scRNA-seq assays across many samples that, for example, represent different ages or healthy and diseased individuals \cite{wagner2018single,stubbington2017single}. On the other hand, the emerging single-cell assays provide a more comprehensive examination of a system, such as single-cell ATAC-seq (scATAC-seq) \cite{buenrostro2015single} that measures chromatin accessibility and single-cell Hi-C \cite{nagano2013single} that explores chromosome architecture.

Integrating the various single-cell assays across different samples provides a comprehensive characterization of a biological system. Many computational methods have been developed to integrate the same single-cell assays of multiple samples \cite{stuart2019integrative,welch2019single,hie2019efficient} or different single-cell assays \cite{forcato2021computational,hao2021integrated}. In the integration of multiple single-cell omics assays, most current methods rely on the known correspondence between features, for example by mapping chromatin loci to genes and assuming the similarity between the samples. The multi-omics integration becomes a harder problem when no prior correspondence is assumed, for example, a gene actually corresponds to multiple loci and accessible loci do not directly indicate gene expression. This leads to a general problem of integrating datasets without known correspondence between features.

When no feature correspondence is given, the structures of the individual datasets can be exploited and matched to integrate the datasets. For example, canonical correlation analysis  examines covariances between the datasets but is limited to deriving linear correspondence between the features. When the datasets are represented as graphs with edges annotating pairs of similar data points within each dataset, the integration problem can be addressed using various graph alignment methods \cite{zhang2020network, o2021graph}. Among the graph alignment methods, \textit{Gromov-Wasserstein optimal transport (GW-OT)} can align graphs based only on the graph structures \cite{memoli2011gromov}. It finds a coupling of the distributions representing the graphs that best preserves the intra-dataset distances between the nodes. 

Optimal transport (OT) compares and finds connections between measures. It seeks the coupling between distributions with the minimum total coupling cost based on predefined costs between locations \cite{Monge_1781,Kantorovitch_1942,Villani_2003}. OT has been a versatile tool widely used in practical problems, such as generative deep learning models \cite{arjovsky2017wasserstein}, domain adaptation \cite{courty2014domain}, and image sciences \cite{ferradans2014regularized}. It has been used to find correspondence between data points in single-cell gene expression data with common features \cite{schiebinger2019optimal,cang2020inferring,nitzan2019gene}. The aforementioned GW-OT has been used in this field to exploit the structural information within individual datasets. SpaOTsc \cite{cang2020inferring} uses fused Wasserstein-Gromov-Wasserstein optimal transport to improve the integration of spatial data and scRNA-seq data with few shared genes by matching the spatial structure and the structure in scRNA-seq data based on gene expression similarity. SCOT \cite{demetci2020gromov} uses Gromov-Wasserstein optimal transport to align scRNA-seq and scATAC-seq data by matching the structures represented by intra-dataset similarity among cells. Pamona \cite{cao2022manifold} uses partial Gromov-Wasserstein optimal transport to partially align scRNA-seq and scATAC-seq data to address the partially overlapping cell populations among different samples.

In addition to studying shared structures revealed by the overlapping part of integrated data, it is equivalently important to examine the structures of non-overlapping parts which may depict a biological process uniquely captured by a certain assay \cite{zhang2021scmc}. Since most integration methods depend on similarities between samples, the dissimilar parts are often overlooked. Efforts have been made to keep the variation among samples examined with the same single-cell assay \cite{zhang2021scmc}.

In the analysis of high-dimensional multimodal datasets, another crucial step is \textit{dimensionality reduction}. Dimensionality reduction is the process of taking high-dimensional data and finding a representation in lower dimensions that is still meaningful. It has many important applications because dimensionality reduction helps address the curse of dimensionality and other challenges that come with working with high-dimensional data \cite{jimenezSupervised}. Principal Component Analysis (PCA) \cite{pearson1901liii} is the most traditional linear technique used in dimensionality reduction but there are many popular non-linear techniques, such as  Local Linear Embedding \cite{roweis2000nonlinear},  Isomap \cite{tenenbaum2000global}, UMAP \cite{mcinnes2018umap},  and t-SNE \cite{van2008visualizing}.

t-SNE is a popular dimensionality reduction and visualization technique that was introduced in 2008 by van der Matten and Hinton \cite{van2008visualizing}. It has been applied to a variety of high dimensional data, including deep learning \cite{Lorincz2021AnOE}, physics \cite{verma2021classification}, and medicine \cite{Abdelmoula12244}. Given a high dimensional dataset, t-SNE outputs a low dimensional representation. t-SNE works by making pairwise affinities between points in high dimensions and pairwise affinities between points in low dimensions. It then uses gradient descent to find the set of points (in low dimensions) that minimize the KL divergence between the two sets of joint probabilities.

In the analysis of multimodal single-cell data, the dimensionality reduction and the integration steps are often performed separately or sequentially, including the existing methods that integrate datasets without known feature correspondences \cite{demetci2020gromov,cao2022manifold}. However, these two steps are closely related in that they both preserve the structures from high dimension to low dimension or from the original spaces to the joint space. The benefit of combining these two steps has been shown in many recent works. For example, MultiMAP performs dimensionality reduction and integration utilizing both shared and non-shared features between datasets \cite{jain2021multimap}. As another example, j-SNE learns a joint representation in low dimensions without shared features across multiple data sets with one-to-one correspondences \cite{do2021generalization}. In this work, we present a workflow called AVIDA (Alternating Method for Visualizing and Integrating Data), that integrates 2D representations of high dimensional data sets by alternating between dimension reduction and alignment. AVIDA operates without knowledge or the necessity of shared features or one-to-one correspondences across data sets. To demonstrate this workflow, we use t-SNE for the dimension reduction module and Gromov-Wasserstein optimal transport for the integration module. {Different choices for the dimension reduction module and alignment module can be utilized in this framework, depending on the application. We also include a small set of additional experiments in \ref{sec:appendix}, which utilize UMAP in the dimension reduction step instead of t-SNE to further demonstrate AVIDA's flexibility as a framework.} In four synthetic datasets and two real biological datasets with ground truth, we show that AVIDA better preserves the structures of the individual datasets while achieving comparable integration quality compared to other methods.

\section{Results}
\label{sec:experiments}
\subsection{Overview of AVIDA}

\begin{figure*}[h]
    \centering
    \includegraphics[width=1.0\textwidth]{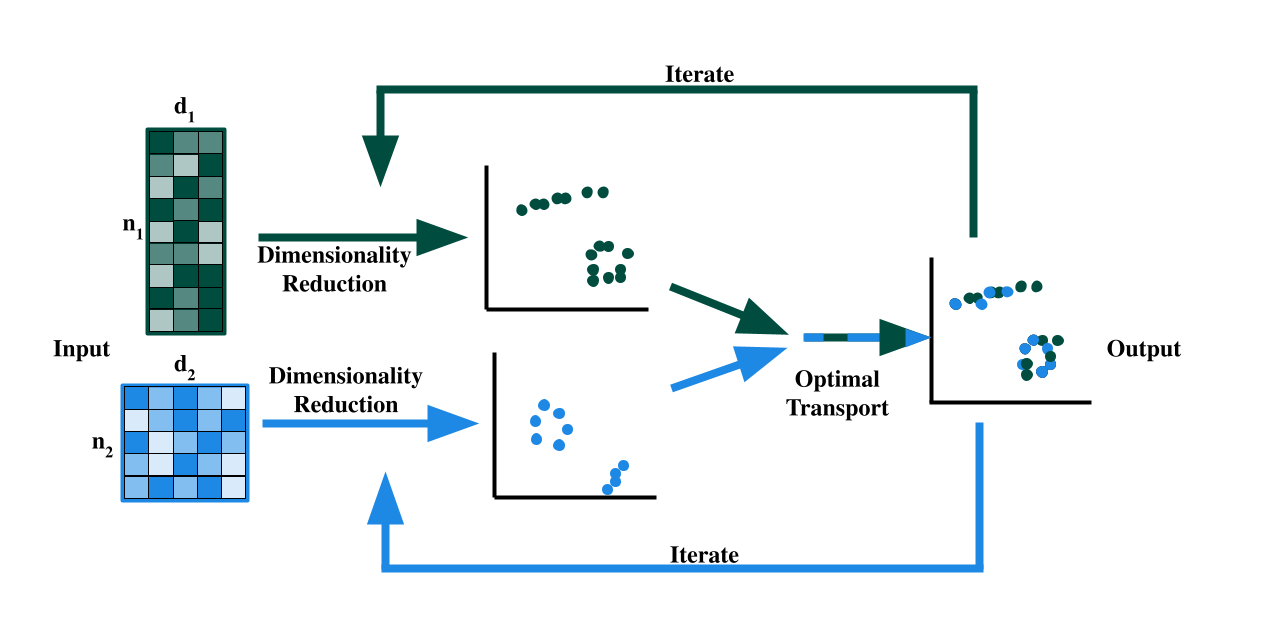}
    \caption{A visual schematic of AVIDA. }
    \label{fig:our_method_diagram}
\end{figure*}

The proposed method is called the \emph{alternating method for visualizing and integrating data}, or AVIDA. AVIDA alternates between improving the low dimensional representation through a dimensionality reduction technique and the alignment of data points in low dimensions across different datasets. The purpose of alternating between dimensionality reduction and alignment is to find a balance between a good representation while still accurately aligning the datasets. We denote AVIDA as a function, taking as input the datasets $X_1, \dots, X_k$, and is parameterized by choice of dimensionality reduction and alignment techniques: $\text{AVIDA}(X_1, X_2, ..., X_k; \text{DR}, \text{ALIGN})$. A simplified schematic of the method is shown in Figure~\ref{fig:our_method_diagram}. As shown in Figure~\ref{fig:our_method_diagram}, AVIDA can take as input two datasets and organizes the data as a pairwise distance matrix. Next, dimensionality reduction using the given pairwise distance matrix is performed on both datasets independently. An alignment method is used to ``align" the datasets in the lower dimensional space and using the aligned data points, a new pairwise distance matrix is formed for each dataset, and the process iterates. This framework is flexible in its choice of dimensionality reduction technique (in fact, different dimension reduction algorithms can be used on different datasets if one so chooses) and alignment method. 

Suppose one is given two datasets $X^{(1)}$ and $X^{(2)}$ and the goal is to create a joint representation of the datasets in a common lower dimensional space.  Using some technique DR for dimensionality reduction (e.g., PCA, t-SNE, Random Forests, etc.) and GW-OT for alignment, we can formulate the objective function for AVIDA as $\text{AVIDA}(X_1, X_2; \text{DR}, \text{GW})$. 
The GW-OT objective is defined with respect to the low dimensional representation of points:
\begin{equation}
    \text{GW}(Y^{(1)},Y^{(2)}) = \sum_{i,j,i^{'},j^{'}} L_{i,j,i',j'}\mathbf{T}_{i,i'}\mathbf{T}_{j,j'} - \epsilon(H(\mathbf{T})), \label{line:gw_loss}
\end{equation}
where $H(\mathbf{T}) = \sum_{i,j} T_{ij} \log(T_{ij})$ is the Entropic regularization term and\newline ${L_{i,j,i',j'}=\|d(y_i^{(1)},y_j^{(1)})-d(y_{i^{'}}^{(2)},y_{j^{'}}^{(2)})\|^2}$ with a chosen distance metric $d(\cdot,\cdot)$. This objective is minimized by using the projected gradient descent method with KL metric-based projections \cite{peyre2016gromov}, ${T\leftarrow \mathrm{Proj}^{\mathrm{KL}}_{U(\mathbf{a},\mathbf{b})}(T\odot e^{-\tau(L\otimes T+\epsilon\log(T))})}$ where\newline  ${U(\mathbf{a},\mathbf{b})=\{T\in\mathbb{R}^{n_1\times n_2}_+: T\mathbbm{1}=\mathbf{a}, T^T\mathbbm{1}=\mathbf{b}\}}$ and $\tau$ is the step size. The implementation in Python Optimal Transport \cite{flamary2021pot} package is used.
The representation for $Y^{(1)}$ will subsequently be mapped to $Y^{(2)}$ using the mapping found by minimizing \eqref{line:gw_loss} with respect to $T$, i.e., by setting $Y^{(1)} = T Y^{(2)}$. Our combined loss function can be represented as
\begin{align}
    &\text{AVIDA}(X^{(1)}, X^{(2)}; \text{DR}, \text{GW})  \nonumber \\& \,\, =  \min_{Y^{(1)},Y^{(2)}} \text{DR}(X^{(1)}, Y^{(1)})+\text{DR}(X^{(2)},Y^{(2)}) + \text{GW}(Y^{(1)},Y^{(2)}), \label{eq:total_loss}
\end{align}
where $\text{DR}(X^{(i)}, Y^{(i)})$ represents the objective loss associated with the dimensionality reduction technique DR. For example, if t-SNE is used for the DR step, the objective can be represented as the KL loss between probability distributions on the points in high and low dimensions. See~\ref{sec:methods} for more details.

\subsection{AVIDA accurately reproduces the intra-dataset structures in integration of synthetic data}
We compared $\text{AVIDA}(X_1, X_2; \text{TSNE}, \text{GW})$ to both Pamona and SCOT across four simulated datasets and two real-world single-cell multi-omics datasets. We chose Pamona and SCOT as a comparison because they are both advanced integration methods {that do not require common features or one-to-one correspondence across data sets}. {To have a fair comparison with SCOT and Pamona, for these experiments we had SCOT and Pamona perform their alignment and then used t-SNE to visualize their low dimensional representations rather than UMAP or PCA. This way we are not comparing different kinds of visualization techniques to each other. To see how these methods would perform using UMAP instead of t-SNE, see Appendix A.} Table~\ref{tab:pamona_met} contains the performance metrics for $\text{AVIDA}(X_1, X_2; \text{TSNE}, \text{GW})$, SCOT and Pamona on both the simulated and real-life datasets. We used five different metrics to assess the performance of these methods: the fraction of samples closer than the true match (FOSCTTM), alignment, integration, accuracy, and representation loss. The accuracy metric is only included on the datasets where the ground truth is known and an empty cell in the table implies the dataset did not meet that requirement. Details on the metrics are included in Section~\ref{sec:metrics}.
 \begin{table*}[h]
\centering
\resizebox{\textwidth}{!}{%
\begin{tabular}{|c|c|c|c|c|c|c|}
\hline
\textbf{Dataset} & \textbf{Method} & \textbf{FOSCTTM} & \textbf{Integration} & \textbf{Accuracy} & \textbf{Alignment}  & \textbf{Representation Loss}  \\ \hline \hline
\multirow{3}{*}{\textbf{Bifurcated Tree}}    & AVIDA      & 0.1202              & 1.0820  & \textbf{4.3863}   & \textbf{0.5157}    & \textbf{0.3275}     \\ \cline{2-7} 
                                   & Pamona               & {\textbf{0.1108}}     &  {\textbf{0.2933}} &  {7.6098} &  {0.9897}    &  {1.0969}     \\ \cline{2-7}
                                   & SCOT                 & 0.2103              & 1.0016 & 12.2095  & 0.75     & 2.1466      \\ \hline \hline
\multirow{3}{*}{\textbf{Circular Frustrum}}    & AVIDA    & 0.1187              & 0.9699 & 2.9377  & \textbf{0.4267}   & \textbf{0.3955} \\ \cline{2-7} 
                                   & Pamona               &  {\textbf{0.0186}}     &  {\textbf{0.2532}} &  {\textbf{1.2577}} &   {0.9363}    &  {0.8377} \\ \cline{2-7}
                                   & SCOT                 & 0.0515              & 1.0032  & 4.3857 & 0.9727    & 1.7083       \\ \hline  \hline
\multirow{3}{*}{\textbf{Dumbbell}}    & AVIDA             & 0.5228              & 0.5568 & 25.1281     & 0.6385   & \textbf{0.1220}        \\ \cline{2-7} 
                                   & Pamona               &  {0.5055}              &  {\textbf{0.3679}} &  {32.1714}   &  {0.7785}  &  {0.6176}              \\ \cline{2-7}
                                   & SCOT                 & \textbf{0.4754}      & 2.565   & \textbf{11.2244}  & \textbf{0.2070}   & 3.6008        \\ \hline   \hline
\multirow{3}{*}{\textbf{Distant Rings}}      & AVIDA      & 0.3138              & 0.6847   & 5.3429  & \textbf{0.639}  & \textbf{0.1916}               \\ \cline{2-7} 
                                   & Pamona               &  {0.2580}              & {1.2407}  &  {1.0}  &  {0.993} &  {1.1784}    \\ \cline{2-7}
                                   & SCOT                 & \textbf{0.0056}     & \textbf{0.0791}  & \textbf{0.2759}    & 0.9125   & 0.9261      \\ \hline  \hline
\multirow{3}{*}{\textbf{sc-GEM}}    & AVIDA               & 0.2070              & 0.4700     & \textbf{2.4996}  & 0.8994 & \textbf{0.4879}    \\ \cline{2-7} 
                                   & Pamona               &  {0.2108}             &  {\textbf{0.3567}} &  {10.894}     &  {0.7237} &  {1.4298}    \\ \cline{2-7}
                                   & SCOT                 & \textbf{0.1818}      & 2.3164 & 6.9267   & \textbf{0.5616}  & 0.8790        \\ \hline  \hline
\multirow{3}{*}{\textbf{scNMT-seq}}    & AVIDA            & 0.2745              & 0.3631  &  4.5787    & \textbf{0.6619 }  & \textbf{1.0489}        \\ \cline{2-7} 
                                   & Pamona             &  {0.3889}             &  {\textbf{0.2446}}  &  {\textbf{0.7032}}     &  {0.9746} &  {4.2435}       \\ \cline{2-7}
                                   & SCOT                 & 0\textbf{.2675}      & 2.4333   & 28.6287    & 0.7522      & 1.1979                \\ \hline 
\end{tabular}%
}
\caption{Metrics for $\text{AVIDA}(X_1, X_2; \text{TSNE}, \text{GW})$ (labeled as AVIDA above), Pamona and SCOT experiments.} 
\label{tab:pamona_met}
\end{table*}

\begin{figure*}[h]
    \centering
    \includegraphics[width=0.9\textwidth]{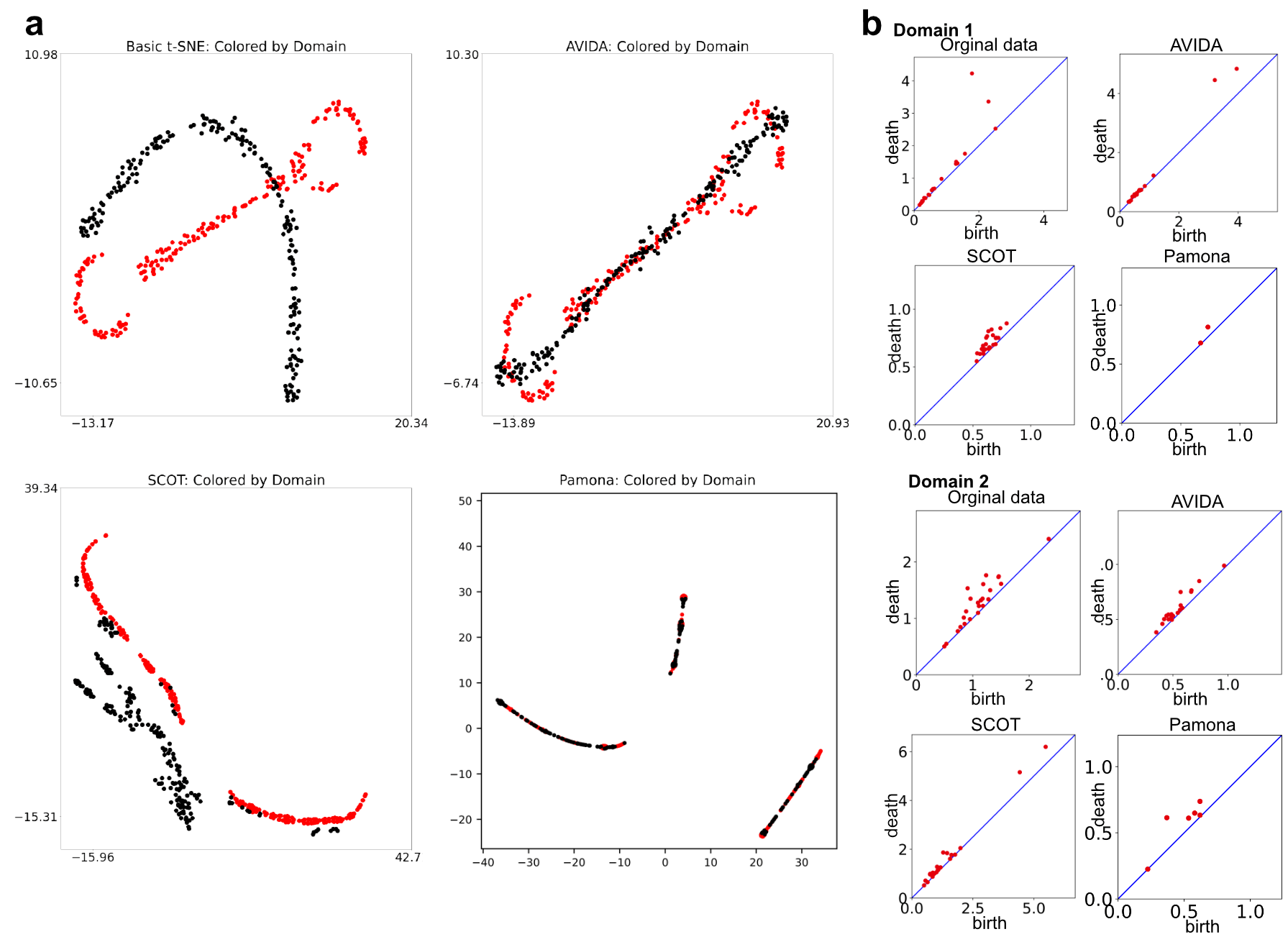}
    \caption{(a) Pamona, AVIDA, SCOT, and t-SNE representation of the dumbbell dataset.  {(b) The $H_1$ persistence diagrams of Vietoris-Rips filtration with Euclidean distance of the original data, and AVIDA and SCOT embeddings. The birth and death values are the scales at which topological features appear and disappear. A point farther away from the diagonal (blue line) represents a significant 1-dimensional loop. ``Domain 1" and ``Domain 2" correspond to the points colored red and black respectively in (a).}}
    \label{fig:dumbbell}
\end{figure*}

 Our four simulated datasets include a bifurcated tree, a circular frustum (from \cite{liu2019jointly}), a dumbbell, and distant rings. The dumbbell and distant rings datasets are introduced in order to highlight the difference between AVIDA and SCOT and Pamona. The dumbbell dataset consists of two rings that are connected by a line. We consider the following split of the dumbbell data set: $X_1$ contains data points from the two rings and a subset of the points along the line connecting the two rings. Then dataset $X_2$ contains all the points along the line connecting the two rings. Thus, the dumbbell dataset allows us to investigate the performance of AVIDA when there is only a partial direct correspondence between data sets.

We also introduce the distant rings dataset. The rings dataset consists of two rings that are far apart from each other in high dimensions. We set the sizes of their radii to be much smaller than the distance between the centers of the rings. Then, the datasets $X_1$ and $X_2$ are generated such that they both contain the entirety of the two rings dataset, i.e. $X_1 = X_2$. This is done so that there is a direct correspondence between points in $X_1$ and $X_2$. Thus, the rings dataset allows us to investigate the performance of AVIDA when there is a full direct correspondence between data sets. In addition, the difference in scale of the distances in the rings dataset allows us to highlight the advantage of using AVIDA rather than other forms of alignment.
    
The specific parameters used to generate these datasets are given in Section~\ref{sec:methods}. The evaluations of these methods on the various metrics are given by Table~\ref{tab:pamona_met}.
 
\begin{figure*}[h]
    \centering
    \includegraphics[width=0.9\textwidth]{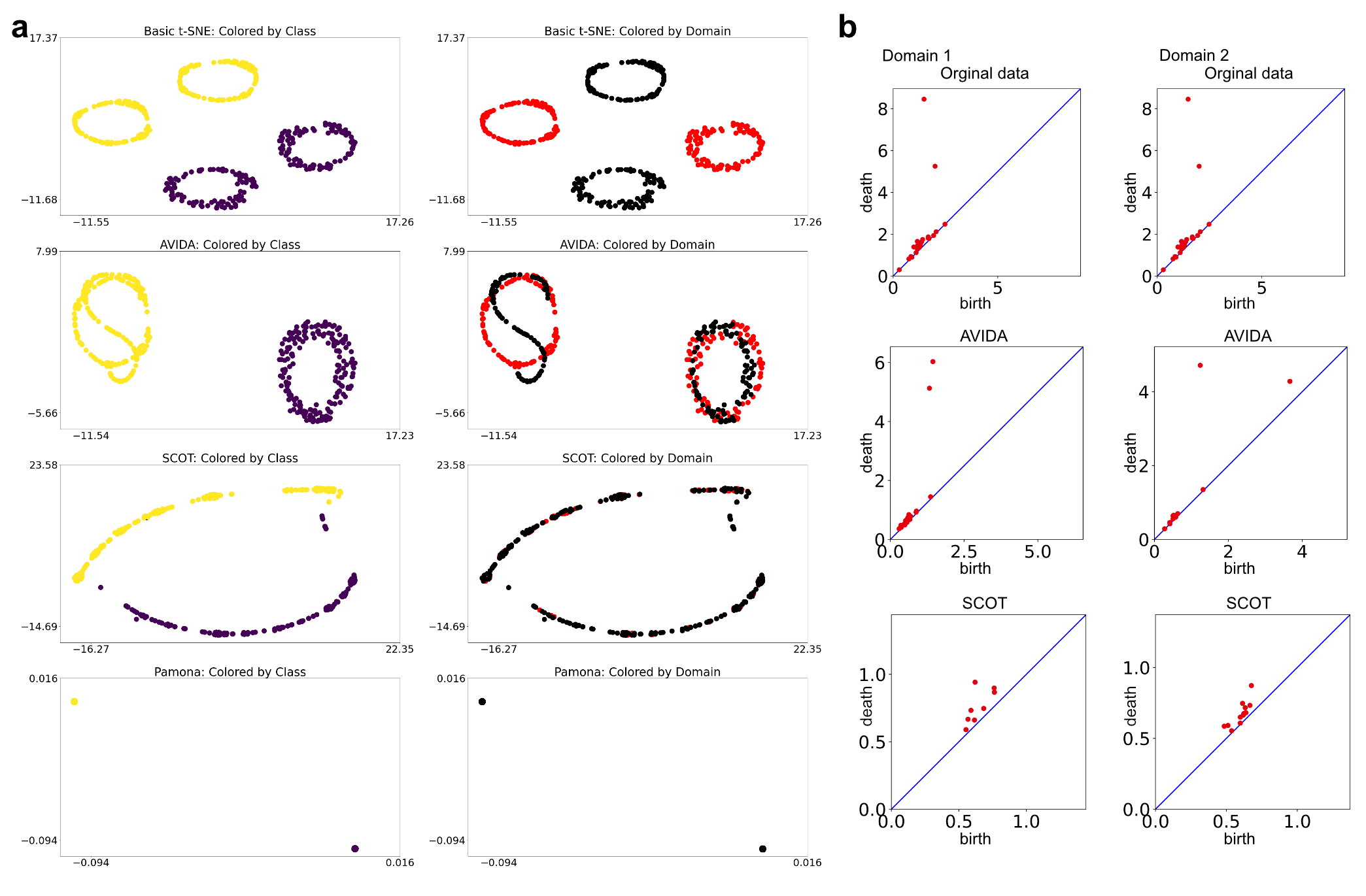}
    \caption{(a) t-SNE, AVIDA, SCOT and Pamona representation of the distant rings dataset.  {(b) The $H_1$ persistence diagrams of Vietoris-Rips filtration with Euclidean distance of the original data, and AVIDA and SCOT embeddings. The birth and death values are the scales at which topological features appear and disappear.} A point farther away from the diagonal (blue line) represents a significant 1-dimensional loop. The $H_1$ diagrams of Pamona embeddings are empty. ``Domain 1" and ``Domain 2" correspond to the points colored red and black respectively in (a). }  
    \label{fig:two_dist_circles}
\end{figure*}

Looking at Figures~\ref{fig:dumbbell} and \ref{fig:two_dist_circles}, it is clear why we want to introduce these datasets. In Figure~\ref{fig:dumbbell}, AVIDA clearly preserved the local structure of both datasets while Pamona and SCOT highlight the linear structure found in both datasets.  {This is demonstrated by both visual inspection of the loop structures preserved by AVIDA, as shown in Figure~\ref{fig:dumbbell}(a) and Figure~\ref{fig:two_dist_circles}(a) and the persistence diagrams, as shown in Figure~\ref{fig:dumbbell}(b) and Figure~\ref{fig:two_dist_circles}(b).} The persistence diagram is the result of persistent homology\cite{edelsbrunner2000topological,zomorodian2004computing} which grows a simplicial complex on a point cloud and tracks the scale at which the topological features appear (birth value) and disappear (death value). A topological feature with large persistence value (difference between birth and death values) is considered significant and we are interested in the one dimensional $H_1$ features that correspond to circles in data. Details of persistent homology are discussed in Section 4.2.2. AVIDA is the only method that is able to successfully integrate the two representations generated by t-SNE's representation. Figure~\ref{fig:two_dist_circles} shows that Pamona's method collapses both rings to a single point, destroying the local structure of the data. SCOT is able to integrate the datasets while still preserving some linear structure but compared to t-SNE's actual 2D representation, AVIDA produces a 2D representation with the most accurate local structure. Since AVIDA allows t-SNE to construct the local structure of the line before mapping, that structure is preserved in the final representation.
 
 However, if we were to look at the FOSCTTM and accuracy scores in Table~\ref{tab:pamona_met} for Figure~\ref{fig:dumbbell} and Figure~\ref{fig:two_dist_circles}, Pamona scores best because all the points are correctly mapped close together. The datasets illustrate our need for a representation metric since the traditional metrics do not penalize for poor representations in 2D. We use t-SNE's loss function as our representation loss since it is a popular dimensionality reduction technique, however, it could easily be replaced by a loss function from other methods (e.g. UMAP).

\subsection{AVIDA achieves a balance between structure representation and multimodal dataset alignment}
We also compare the outputs from two real-world single-cell multi-omics datasets. The first is sc-GEM, a dataset from \cite{cheow2016single} which contains both gene expression and DNA methylation at multiple loci on human somatic cell samples under coversion to induced pluripotent stem cells. The second is scNMT-seq, a dataset of chromatin accessibility, DNA methylation, and gene expression on mouse gastrulation samples collected at four different time states from \cite{argelaguet2019multi}. The evaluations of AVIDA, SCOT, and Pamona on these datasets are also given in Table~\ref{tab:pamona_met}. 
\begin{figure}[h]
    \centering
    \includegraphics[width=0.45\textwidth]{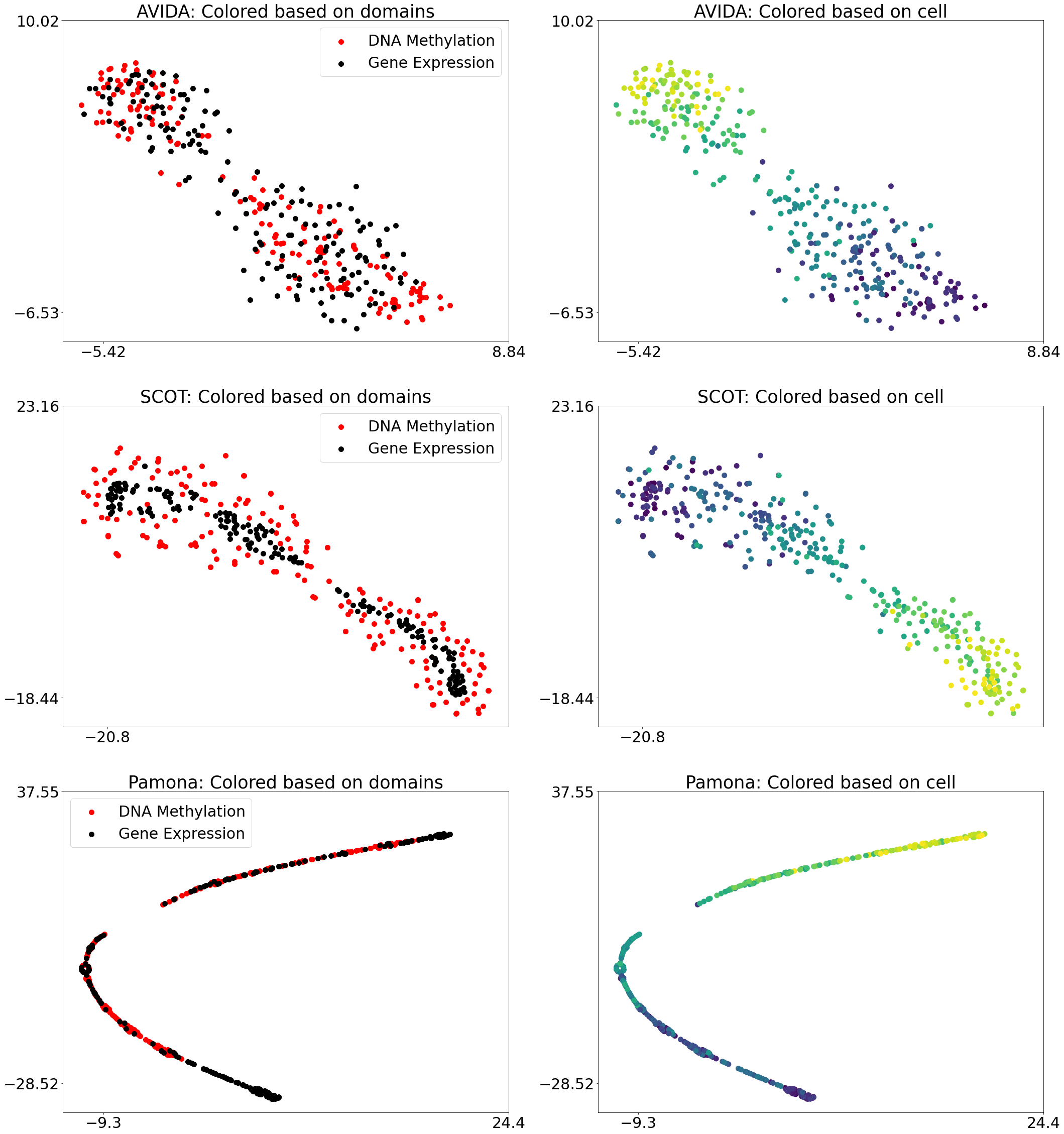}
    \caption{AVIDA, SCOT and Pamona representation of sc-GEM. {The visualizations for each of the methods were made by t-SNE.}}
    \label{fig:scGEM}
\end{figure}
In Figure~\ref{fig:scGEM}, we can see the different 2D representations for sc-GEM. The left column of the figure shows the integration between the two datasets and the right column has the data points colored by cell. From these representations, we can see that AVIDA is able to fully integrate the two different datasets where there is some noticeable separation in the SCOT representation. Since this dataset contains the conversion from somatic cells to stem cells,  {we hope to see a gradient of colors from one end of the representation to the other which all methods are able to achieve. This is a good example of how AVIDA's performance on integration of real-life datasets is comparable to both SCOT and Pamona.}

We can also confirm this observation in Table~\ref{tab:pamona_met}. AVIDA is able to achieve FOSCTTM and alignment scores that are comparable to SCOT and Pamona while simultaneously having the best representation loss. The same holds true for scNMT-seq as well. These examples illustrate that AVIDA is comparable to both Pamona and SCOT on real-life datasets while also performing well on the adversarial datasets: the dumbbell and distant rings datasets.

\begin{figure}[h]
    \centering
    \includegraphics[width=0.45\textwidth]{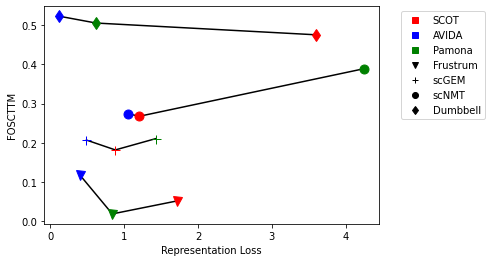}
    \caption{A comparison of methods using integration and 2D representation.}
    \label{fig:vis_vs_int}
\end{figure}

While we did not plot every dataset's low dimensional representation here, Figure~\ref{fig:vis_vs_int} compares the FOSCTTM and representation losses for each 2D representation generated by SCOT, AVIDA, and Pamona. The shapes designate the dataset's low dimensional representation and the different colors represent the method that was used. We can see that across the different datasets, all three methods have comparable FOSCTTM scores, indicating that the integration of the datasets are similar. However, we can also see that AVIDA by far has the best representation loss, indicating a more accurate low dimensional representation.

\section{Discussion}

Motivated by the similar fundamental assumptions in dimension reduction and data integration that they both try to preserve the structures of datasets, we developed an alternating method, AVIDA, which combines these two processes for joint 2D representation of datasets without shared features. Comparing with the methods that perform integration first and then dimension reduction, AVIDA better preserves the detailed structures of the datasets being integrated especially the structures present in only one of the datasets. This property allows the identification of mechanisms that can only be revealed with one of the technologies.

In this work, we demonstrate the method using t-SNE for dimension reduction and Gromov-Wasserstein optimal transport for data integration. In general, other dimension reduction methods and integration methods could be used. The representation loss used in the comparison can also be used as a control metric about how well the structures of individual datasets are preserved in the joint representation. This metric can be used to find a balance between integration and representation when other methods are used for the dimension reduction and integration modules. The comparison indicates that a method could do a perfect job in integration while missing structures presented in the individual datasets. It is thus important to also evaluate the quality of the structure representation of individual datasets when developing joint dimension reduction methods for high-dimensional multimodal datasets.

Despite the improvements on performing the two processes separately, the quality of the joint 2D representation still heavily depends on the performance of the specific dimension reduction method and integration method. While the quality of dimension reduction can be checked by comparing it to the structures present in the original high dimensional datasets, it is hard to evaluate the integration quality without ground truth. It is thus also important to further validate the result with prior knowledge or assess the robustness of the integration with, for example, subsampling.

Upon the joint representation of multimodal datasets, one major downstream task is to find the correspondence between the non-overlapping features across the datasets. A potential method for this is to track the contributions of original features to the common low dimensional representations and subsequently find the correspondence between them.

\section{Methods}
\label{sec:methods}

AVIDA is a framework that takes input data sets $\{ X^{(\ell)} \}_{i=1}^N$ where the data sets $X^{(\ell)} \in \mathbb{R}^{n_\ell \times d_\ell}$ need not be in the same feature space. The output of AVIDA is a low dimensional representation of all data sets simultaneously in a single feature space. This is accomplished by alternating between dimensionality reduction and alignment. The AVIDA framework is presented in Algorithm~\ref{alg:our_method_frame}. The choice of dimensionality reduction technique and alignment method is up to the user and can be chosen based on the use case. In Section~\ref{sec:tavida}, we present a detailed implementation of AVIDA using t-SNE for dimensionality reduction and GW-OT for alignment. 

\begin{algorithm}[th]
\caption{AVIDA}\label{alg:our_method_frame}
\begin{algorithmic}
    \State \textbf{Input}:$N$ datasets $X^{(\ell)} = \{x_i^{(\ell)} \}_{i=1}^{n_\ell} \subset \mathbb{R}^{d_\ell}$, target dimension $d$, Dimensionality Reduction Method $DR(\cdot)$, Alignment Method $ALIGN(\cdot)$.
    \State{\textbf{Output}: Low-dimensional representations $Y^{(\ell)} = \{ y_i^{(\ell)}\}_{i=1}^{n_\ell} \subset \mathbb{R}^{d}$.}
    \State Initialize $Y^{(\ell)}_0$ for $\ell \in [N]$ and set $t=0$.
    \Do
    \State Dimensionality reduction step:
    \State \indent  $\hat{Y}^{(\ell)}_t = DR(X^{(\ell)},Y^{(\ell)}_t)$ for $\ell \in [N]$.
    \Comment{Input dataset $X^{(\ell)}$ and initialization $Y^{(\ell)}_t$}
    \State Alignment step:
    \State \indent $[Y^{(1)}_{t+1}, \cdots, Y^{(N)}_{t+1}] = ALIGN(\hat{Y}^{(1)}_{t}, \cdots, \hat{Y}^{(N)}_{t})$.
    \State Increment iteration count: $t = t+1$.
    \doWhile{stopping criteria not satisfied}
    \State Return $Y^{(\ell)} = Y^{(\ell)}_{t}$ for $\ell \in [N]$.
    \end{algorithmic}
\end{algorithm}

\subsection{AVIDA with t-SNE and GW-OT}
\label{sec:tavida}
In this section, we present our implementation of the AVIDA framework using t-SNE for dimensionality reduction and GW-OT for alignment, i.e., $\text{AVIDA}(X_1, X_2; \text{TSNE}, \text{GW})$. For simplicity, we assume there are two input data sets $X^{(1)} = \{x_i^{(1)}\}_{i=1}^{n_1} \subset \mathbb{R}^{d_1}$ and $X^{(2)} = \{x_i^{(2)}\}_{i=1}^{n_2}\subset \mathbb{R}^{d_2}$ and that the low dimensional output feature space has dimension $d=2$, i.e., $Y^{(1)} =  \{y_i^{(1)}\}_{i=1}^{n_1} \subset \mathbb{R}^{2}$ and $Y^{(2)} = \{y_i^{(2)}\}_{i=1}^{n_2} \subset \mathbb{R}^{2}$.

In the dimensionality reduction step, t-SNE generates pairwise affinity values $\{p_{ij}^{(\ell)}\}$ for each of the dataset $X^{(\ell)}$, as given by 

\begin{equation}
    p_{j\vert i}^{(\ell)} = \frac{\exp(-\|x_i^{(\ell)}-x_j^{(\ell)}\|^2/2\sigma_i^{(\ell)})}{\sum_{k\neq i} \exp(-\|x_k^{(\ell)}-x_i^{(\ell)}\|^2/2\sigma_i^{(\ell)})} \label{line:pj_i_value}
\end{equation}
\begin{equation}
        p_{ij}^{(\ell)} = \frac{p_{j\vert i}^{(\ell)}+p_{i\vert j}^{(\ell)}}{2n_\ell}, \label{line:pij_value}
\end{equation}
where the $\sigma_i^{(\ell)}$'s satisfy
\begin{equation}
    \rho = 2^{-\sum_{j\neq i} p_{j\vert i}^{(\ell)}\log(p_{j\vert i}^{(\ell)})},
\end{equation}
for a perplexity value $\rho$ chosen by the user. To obtain $y_i^{(\ell)}$, t-SNE minimizes the Kullback-Leibler divergence between $\{p_{ij}^{(\ell)}\}_{j\neq i}$ and $\{q_{ij}^{(\ell)}\}_{j\neq i}$ using gradient descent. The target probabilities $q_{ij}^{(\ell)}$ are defined as:
\begin{equation}
        q_{ij}^{(\ell)} = \frac{(1+\|y_i^{(\ell)}-y_j^{(\ell)}\|^2)^{-1}}{\sum_{i^{'},j^{'}} (1+\|y_{i^{'}}^{(\ell)}-y_{j^{'}}^{(\ell)}\|^2)^{-1}}. \label{line:qij}
\end{equation}
To obtain $y_i^{(\ell)}$, t-SNE minimizes the Kullback-Leibler divergence between $\{p_{ij}^{(\ell)}\}_{j\neq i}$ and $\{q_{ij}^{(\ell)}\}_{j\neq i}$ using gradient descent: 
\begin{equation}
    KL(P_\ell \vert \vert Q_\ell) = \sum_{i,j=1}^{n_\ell} p_{ij}^{(\ell)}\log\left(\frac{p_{ij}^{(\ell)}}{q_{ij}^{(\ell)}}\right), \label{eq:tsne_loss_sec4}
\end{equation}
The t-SNE method utilizes a ``early exaggeration" phase to artificially highlights the attractions between points in similar neighborhoods, promoting clusters. This period is a very important tool that allows t-SNE to develop local structures in its representation. The early exaggeration phase occurs in the first 200 iterations of gradient descent in which $p_{ij}^{(\ell)}$ values are scaled by a factor of 4. It has been shown that the early exaggeration phase in t-SNE promotes clustering of similar points \cite{linderman2019clustering}. After the first 200 iterations, the $p_{ij}^{(\ell)}$ values are returned to their original value and t-SNE continues to perform gradient descent. 

In the alignment step of AVIDA, GW-OT is used to align data points across data sets. Given the current low dimensional representations outputs from t-SNE, $Y^{(1)}$ and $Y^{(2)}$, the following optimization problem is solved to compute the transport matrix $\mathbf{T}$:
\begin{align}
    &\text{GW}(Y^{(1)},Y^{(2)}) \nonumber\\ & \,\, = \min_{\mathbf{T}}  \sum_{i,j,i^{'},j^{'}} \|d(y_i^{(1)},y_j^{(1)})-d(y_{i^{'}}^{(2)},y_{j^{'}}^{(2)})\|^2\mathbf{T}_{i,i^{'}}\mathbf{T}_{j,j^{'}} - \epsilon(H(\mathbf{T})), \label{line:gw_loss_sec4}
\end{align}
where $H(\mathbf{T}) = \sum_{i,j} \mathbf{T}_{ij} \log(\mathbf{T}_{ij})$ is an Entropic regularization term and $d(\cdot,\cdot)$ is a chosen distance metric. The representation for $Y^{(1)}$ is mapped to $Y^{(2)}$ using the mapping found by minimizing \eqref{line:gw_loss_sec4}, or by computing $Y^{(1)} = T Y^{(2)}$. $\text{AVIDA}(X^{(1)}, X^{(2)}; \text{TSNE}, \text{GW})$ continues alternating between minimizing the KL loss in t-SNE and using optimal transport to align points until a stopping criteria is reached. In this implementation, we choose to limit the number of iterations to 1000 and perform alignment every 100 iterations after the early exaggeration phase (i.e., after the first 200 iterations) of t-SNE. The pseudo-code for $\text{AVIDA}(X^{(1)}, X^{(2)}; \text{TSNE}, \text{GW})$ is provided in Algorithm~\ref{alg:our_method}.

\begin{algorithm}[!h]
\caption{$\text{AVIDA}(X_1, X_2; \text{TSNE}, \text{GW})$}\label{alg:our_method}
\begin{algorithmic}
    \State \textbf{Input}: datasets $X^{(1)} = \{x_1^{(1)},\ldots,x_{n_1}^{(1)}\}$, $X^{(2)} = \{x_1^{(2)},\ldots,x_{n_2}^{(2)}\}$, perplexity $\rho$, and regularization parameter $\epsilon$
    \State{\textbf{Output}: low-dimensional representations: $Y^{(1)}_0=\{y_1^{(1)},\ldots,y_{n_1}^{(1)}\}$,  $Y^{(2)}_0=\{y_1^{(2)},\ldots,y_{n_2}^{(2)}\}$}
    \State{Compute pairwise affinities $p_{ij}^{(1)}$, $p_{ij}^{(2)}$ with perplexity $\rho$ (using Eq. \eqref{line:pj_i_value} and Eq.~ \eqref{line:pij_value})}
    \State{Initialize solutions $Y^{(1)}_0,Y^{(2)}_0$ with points drawn i.i.d. from $\mathcal{N}$(0,$10^{-4}I$)}
    \While{$t < 1000$}
    \If{$\mod(t, 100) \neq 0$} \For{$\ell = {1,2}$}
    \State{Compute pairwise affinities $q_{ij}^{(\ell)}$ (using Eq. \ref{line:qij})}
    \State{Compute gradients $\Delta_t^{(\ell)} = \frac{\delta }{\delta {Y_{t}^{(\ell)}}} \text{TSNE}(X^{(\ell)}, Y^{(\ell)}_t)$ (using Eq.~\ref{eq:tsne_loss_sec4})}
    \State{Set $Y_t^{(\ell)} = Y_t^{(\ell)} + \Delta_t^{(\ell)}$}
    \EndFor
    \Else
    \State{Compute the GW-OT mapping, $\mathbf{T}$, between $Y_t^{(1)}$ and $Y_t^{(2)}$ (using Eq. \ref{line:gw_loss})}
    \State{Set $Y_{(t+1)}^{(\ell)} = $$\mathbf{T}$ $Y_t^{(\ell)}$}
    \EndIf
    \State{$t \gets t+1$}
    \EndWhile
    \end{algorithmic}
\end{algorithm}

\subsection{Metrics, parameters, hardware}
\label{sec:metrics}
The metrics used in Section~\ref{sec:experiments} are described in detail in this section. For reproducibility, we also include the hardware settings under which these experiments were run and the user-selected parameters employed to obtain our numerical results.\\
\subsubsection{Metrics}
To compare $\text{AVIDA}(X_1, X_2; \text{TSNE}, \text{GW})$, Pamona, and SCOT five different metrics are employed: fraction of samples closer than the true match (FOSCTTM), alignment, integration, accuracy, and representation loss. The FOSCTTM and alignment are metrics proposed in previous works. FOSCTTM was originally proposed by Liu et al. \cite{liu2019jointly} and was used to validate the performance of SCOT. The alignment score was used in \cite{cao2022manifold} to compare Pamona and SCOT. In addition to the metrics used in previous works, we also introduce a few others to capture various aspects of the output representation. The additional metrics we measure are integration, accuracy, and representation loss. In this section, we define each and the conditions under which these metrics are meaningful. For notational simplicity, $D \in \mathbb{R}^{n_1\times n_2}$ such that $D_{ij} = d(y^{(1)}_{i}, y^{(2)}_{j})$ denote the pairwise distance matrix between points in $Y^{(1)}$ and points in $Y^{(2)}$.

The FOSCTTM captures roughly the accuracy of the representation. FOSCTTM operates under the assumption that every point has a ``true match" and that the ``true matches" should be close together in the lower dimensional representation. This is formalized as follows. Assume, for simplicity, and $n_1 = n_2 = n$ and without loss of generality that the true match of $x^{(1)}_i$ is $x^{(2)}_i$ for all $i \in [n]$. The FOSCTTM is defined as:
        \begin{equation}
           \text{FOSCTTM} = \sum_{i=1}^{n} \frac{\vert \{j: D_{ij} < D_{ii}  \} \vert }{n-1} +  \sum_{j=1}^{n} \frac{\vert \{i: D_{ij} < D_{jj}  \} \vert }{n-1}. 
           \label{eq:foscttm}
        \end{equation}
In other words, for each point $Y^{(1)}$, determine the fraction of the points $y_i^{(2)}$ that are closer to $y_i^{(1)}$ than $y_i^{(2)}$. Then, repeat the process for points in $Y^{(2)}$. Smaller values of FOSCTTM indicate better performance.

Under these same assumptions (that every point has a true match), we can also define an accuracy score. The idea is that points that are true matches should appear close together in the lower dimensional representation. This is measured by taking a simple trace of the matrix $D$:
$$\text{Accuracy} = \sum_{i=1}^n D_{ii} = tr(D)$$

The Alignment score used in this work was also used in \cite{cao2022manifold}. The alignment score measures how well aligned the two datasets being integrated are in low dimensions. For the alignment score, we assume that each data set has class labels and that those class labels can be shared across data sets. The points in each data set are split into ``shared" and ``dataset specific". ``Shared" data points have representation in both $Y^{(1)}$ and $Y^{(2)}$ whereas ``dataset specific" data points only appear in one of the datasets. The alignment score is computed as follows. Let $S^{(1)} \cup P^{(1)} = Y^{(1)}$ and $S^{(2)} \cup V^{(2)} = Y^{(2)}$ where sets $S^{(\ell)}$ denote the set of all points corresponding to ``shared" data points and $V^{(\ell)}$ denote the set indices of all dataset specific points in $Y^{(\ell)}$. The alignment score is defined as:
$$ \text{Alignment} = 1 - \frac{\vert\bar{x}_s - k/{(\ell+1)}\vert }{k - k/{(\ell+1)}},$$
where $\bar{x}_s$ is the average number of nearest neighbors that are shared points from the same dataset.

The aforementioned metrics have been utilized in previous works. We also propose to use the following for evaluating the representation of the low dimensional data. First, we employ a symmetrized Kullback-Leibler loss with a student t-distribution kernel to evaluate how well the output represents the high dimensional data in an integrated fashion. We refer to this as the Representation Loss:
\begin{align*}
\text{Representation Loss} &= \frac{1}{2} \left(\text{KL}(X^{(1)}\| Y^{(1)}) +  \text{KL}(Y^{(1)}\| X^{(1)}) \right) \\& + \frac{1}{2} \left(\text{KL}(X^{(2)}\| Y^{(2)})+\text{KL}(Y^{(2)}\| X^{(2)})\right). 
\end{align*}
 {The choice of this representation loss as a way to measure the quality of the representation in 2D is based on the fact that popular data dimensionality reduction techniques such as UMAP and t-SNE, both use a version of the KL loss. We recognize that there are other dimensionality reduction techniques, such as PCA or Laplacian Eigenmaps. However such techniques are spectral methods whose loss functions are evaluated by manifold-based metrics similar to FOSCTTM ~\eqref{eq:foscttm} and Integration~\eqref{eq:integration}. This representation loss is a way to measure the quality of the representation in cases of structures that are not best described by the alignment of nearest neighbors, such as clusters or rings. Since t-SNE and UMAP are most adept at preserving these structures in low dimensions, it seems natural to modify their loss function as a way to measure the quality of the 2D representations.}

Lastly, we want to evaluate how well integrated the two data sets are in low dimensions. We say that integration is the average, minimum distance between a data point in $Y_1$ and any data point in $Y_2$. The integration is defined as: 
\begin{equation}
    \text{Integration} = \frac{1}{n_1} \sum_{i=1}^{n_1} \min_j D_{ij} + \frac{1}{n_2} \sum_{j=1}^{n_2} \min_i D_{ij}.
    \label{eq:integration}
\end{equation} 
    

 {
\subsubsection{Persistent homology}
Persistent homology \cite{edelsbrunner2000topological,zomorodian2004computing} is used to evaluate the conservation of local geometries of the synthetic datasets. On a point cloud, a filtration of a simplicial complex $K$ such that $\emptyset = K^0\subset K^1\subset \dots \subset K^m=K$ is constructed based on certain rules such as the Vietoris-Rips filtration, which we employ here. For each simplicial complex $K^i$, the rank of the $k$th homology group $H_k(K^i)$ represents the $k$th Betti number of $K^i$. For the examples here, we focus on the $1$st homology group which represents the $1$-dimensional holes in the data such as loops and rings. Along the filtration, the appearance and disappearance of these homology groups are tracked by computing the $p$-persistent $k$th homology group of $K^i$, $H_k^p(K^i)$ which records the homology classes of $K^i$ that persist at least until $K^{i+p}$. Each homology class is then represented by a pair of filtration values at which the class appears and disappears, usually called the birth and death values. These outputs of persistent homology can be visualized as persistence diagrams by taking the birth and death values as 2D coordinates. A more persistent homology class (with a large difference between death and birth values or equivalently farther away from the diagonal in the persistence diagram plots) is considered a significant feature. For the examples here, we are interested in the significant $1$-dimensional loops which are captured as significant off-diagonal points in the $H_1$ persistence diagram. We refer interested readers to \cite{edelsbrunner2022computational} for complete details of persistent homology. Here, the package Dionysus 2 \cite{dionysus2} was used for persistent homology computation with Vietoris-Rips filtration on Euclidean distance.
}

\subsubsection{Parameters}
The default perplexity value in most standard implementations of t-SNE is 30. However, depending on the dataset, the perplexity value may need to be adjusted. Table~\ref{tab:perplexity} shows the perplexity value choices for each experiment presented in Section~\ref{sec:experiments}.
 \begin{table*}[h]
\centering
\resizebox{\textwidth}{!}{%
\begin{tabular}{|c|c|c|c|c|c|c|}
\hline
\textbf{Dataset} &\textbf{Bifurcated Tree} & \textbf{Circular Frustrum} & \textbf{Dumbbell} & \textbf{Distant Rings} & \textbf{sc-GEM} & \textbf{scNMT-seq}  \\ \hline
\textbf{Perplexity Value} & 30 & 60 & 30 & 30 & 50 & 100 \\ \hline
\end{tabular}%
}
\caption{Perplexity choices for each dataset.}
\label{tab:perplexity}
\end{table*}
In addition to perplexity, another important parameter is $\varepsilon$ in Equation \ref{line:gw_loss}. For all of our experiments, $\varepsilon$ was set to be $5 \times 10^{-3}$ but depending on the dataset could be adjusted.
\subsubsection{Hardware}
We ran the experiments on an Intel i7-10750H CPU (base frequency 2.60GHz) with 8GB memory.
\subsection{Datasets}
\label{sec:data}
For our analysis, we introduced two synthetic datasets: the dumbbell dataset and distant rings dataset. The dumbbell dataset consists of two sub-datasets, $X^{(d,1)}, X^{(d,2)} \subset \mathbb{R}^{2}$ with 200 datapoints each. For all $0 \leq i \leq 200$,
\begin{align*}
    X_{i,1}^{(d,1)} &\sim 50U(0,1) \\
    X_{i,2}^{(d,1)} &\sim N(0,1)
\end{align*}
where $U(0,1)$ is the uniform distribution and $N(0,1)$ is the normal distribution. This essentially constructs $X^{(d,1)}$ as a line in 2D with a little bit of noise. To construct the two rings in $X^{(d,2)}$, we consider $\theta \sim U(0,2\pi)$ and $r \sim N(3,0.5)$, then use it in our construction.
\begin{align*}
    X_{i,1}^{(d,2)} &\sim  r\cos(\theta), \hspace{5mm} 1 \leq i \leq  50 \\
    X_{i,2}^{(d,2)} &\sim r\sin(\theta),\hspace{5mm} 1 \leq i \leq  50\\
    X_{i,1}^{(d,2)} &\sim r\cos(\theta) + 14,\hspace{5mm} 50 < i \leq 100 \\
    X_{i,2}^{(d,2)} &\sim r\sin(\theta),\hspace{5mm} 50 < i \leq 100\\
\end{align*}
The first 50 points in $X^{(2)}$ are a slightly noisy circle centered at 0, where the next 50 points in the dataset are the same slightly noisy circle centered instead at 14. These two rings are then connected by a line.
\begin{align*}
    X_{i,1}^{(d,2)} &\sim U(3,10),\hspace{5mm} 100 < i \leq 200 \\
    X_{i,2}^{(d,2)} &\sim N(0,0.2),\hspace{5mm} 100 < i \leq 200
\end{align*}
This line is the last 100 points and also has small noise across one dimension.

The distant rings dataset also contains two subdatasets, $X^{(c,1)},X^{(c,2)}\subset \mathbb{R}$. Again, we let $\theta \sim U(0,2\pi)$ and now we define $r_1 \sim N(5,1)$ and $r_2 \sim N(5,0.1)$ and define two different rings.
\begin{align*}
    X_{:,1}^{(c,1)} &\sim r_1\cos(\theta) \hspace{5mm} \\
    X_{:,2}^{(c,1)} &\sim r_1\sin(\theta) \hspace{5mm} \\
    X_{:,1}^{(c,2)} &\sim r_2\cos(\theta) + 100 \hspace{5mm} \\
    X_{:,2}^{(c,2)} &\sim r_2\sin(\theta) + 100 \hspace{5mm}
\end{align*}
Essentially for each dataset, we construct two rings where the distance between them dwarfs the radius of each ring. To make these two rings distinct, we constructed one ring to have much less noise than the other. 
\section{Data Availability}
The synthetic data distant rings and dumbbell dataset are available at \url{https://github.com/kat-dover/AVIDA/tree/main/data} and the bifurcated tree and circular frustum were downloaded from the SCOT repository \url{https://rsinghlab.github.io/SCOT/data/}. The sc-GEM data from \cite{cheow2016single} was downloaded from the SCOT repository given at \url{https://rsinghlab.github.io/SCOT/data/}.  The scNMT-seq data from  \cite{argelaguet2019multi} were downloaded from the Pamona repository given at \url{https://github.com/caokai1073/Pamona}.

\section{Code Availability}
The AVIDA implementation with t-SNE as the dimension reduction module and Gromov-Wasserstein optimal transport as the alignment module is available at \url{https://github.com/kat-dover/AVIDA} which will be made publically available on Github upon publication.

\section*{Acknowledgement}
R.V. acknowledges support from NSF DMS-1954233, NSF DMS-2027299, U.S.
Army 76649-CS, and NSF+Simons Research Collaborations on the
Mathematical and Scientific Foundations of Deep Learning. Q.N. was partly supported by a NSF grant DMS-1763272 and a grant from the Simons Foundation (594598, Q.N.). Z.C. was partly supported by a start-up grant of North Carolina State University and NSF grant DMS-2151934. 
 \bibliographystyle{elsarticle-num} 
 \bibliography{cas-refs}





\appendix
\section{Using Alternate Dimensionality Reduction Techniques}
\label{sec:appendix}
 {We introduce AVIDA as a framework that allows for different methods for dimension reduction (or visualization) and alignment can be used depending on the dataset and applications. UMAP is another common dimensionality reduction technique utilized in computational biology. Here, we demonstrate AVIDA using UMAP for the dimension reduction module and GW-OT for the alignment module. The purpose of these brief experiments is to demonstrate AVIDA's viability as a framework. The experiments here essentially replicate a small subset of the experiments presented in the main section of our paper with the main difference being the utilization of UMAP for dimension reduction instead of t-SNE. To create 2D representations for SCOT and Pamona, we also used UMAP. }

\begin{figure*}[h]
    \centering
    \includegraphics[width=\textwidth]{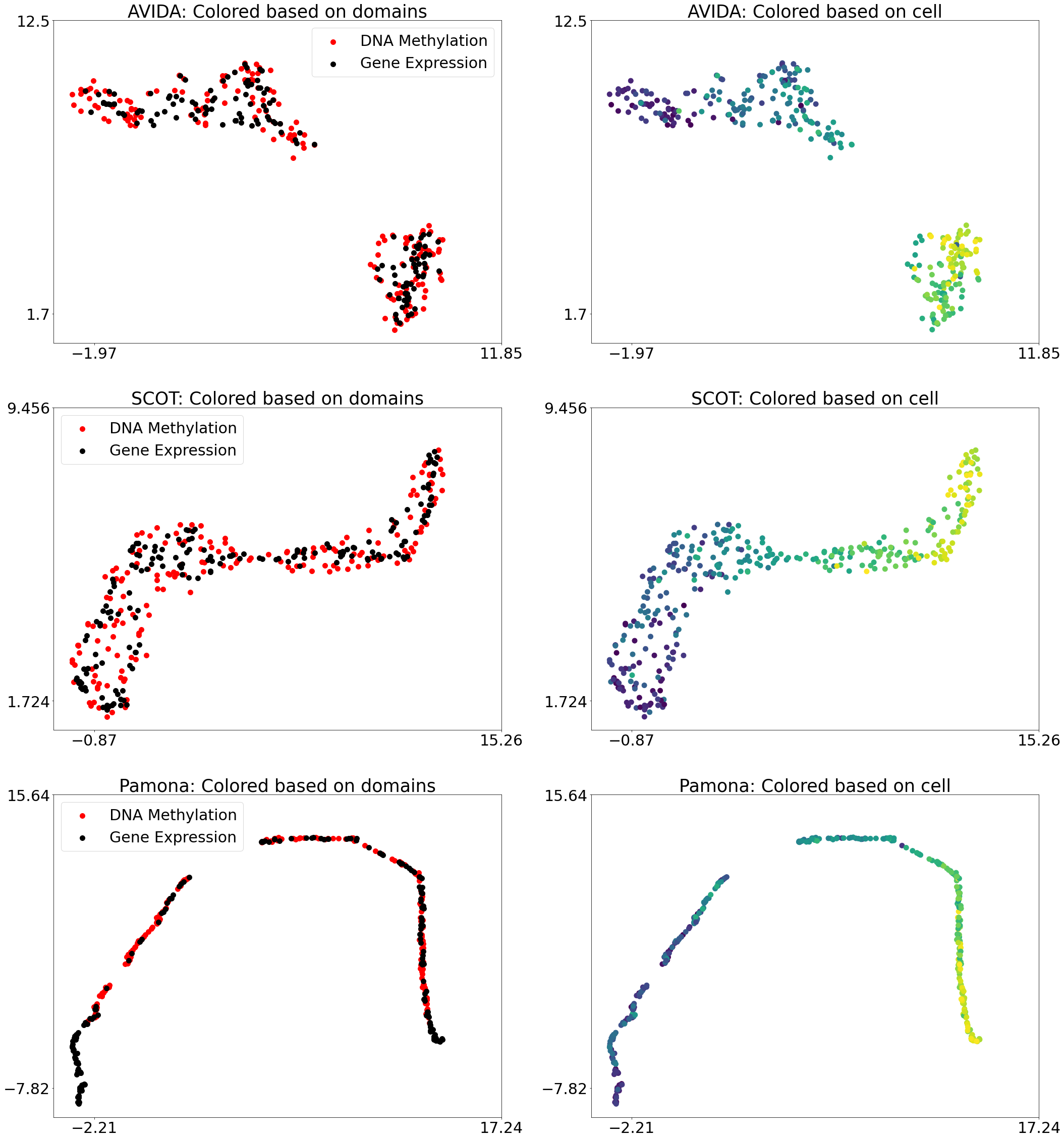}
    \caption{ AVIDA, SCOT, and Pamona representation of the scGEM dataset. In this experiment, UMAP was applied to SCOT and Pamona's output and UMAP's gradient was incorporated into AVIDA for the dimension reduction module. }
    \label{fig:umap_scGEM}
\end{figure*}

 {
In Figure~\ref{fig:umap_scGEM}, we apply $\text{AVIDA}(X_1, X_2; \text{UMAP}, \text{GW})$ to the sc-GEM dataset, a dataset from \cite{cheow2016single} which contains both gene expression and DNA methylation at multiple loci on human somatic cell samples under coversion to induced pluripotent stem cells.
We can see comparing Figure~\ref{fig:umap_scGEM} (which uses UMAP for dimension reduction) with Figure~\ref{fig:scGEM} (which uses t-SNE for dimension reduction), using UMAP produces nearly the same clusters, but here we see a more distinct separation between the two point clouds, both for AVIDA and for Pamona. This shows that there may be datasets where another dimensionality reduction technique might be superior over other choices. However, the reverse can also be true. }

\begin{figure*}
    \centering
    \includegraphics[width=\textwidth]{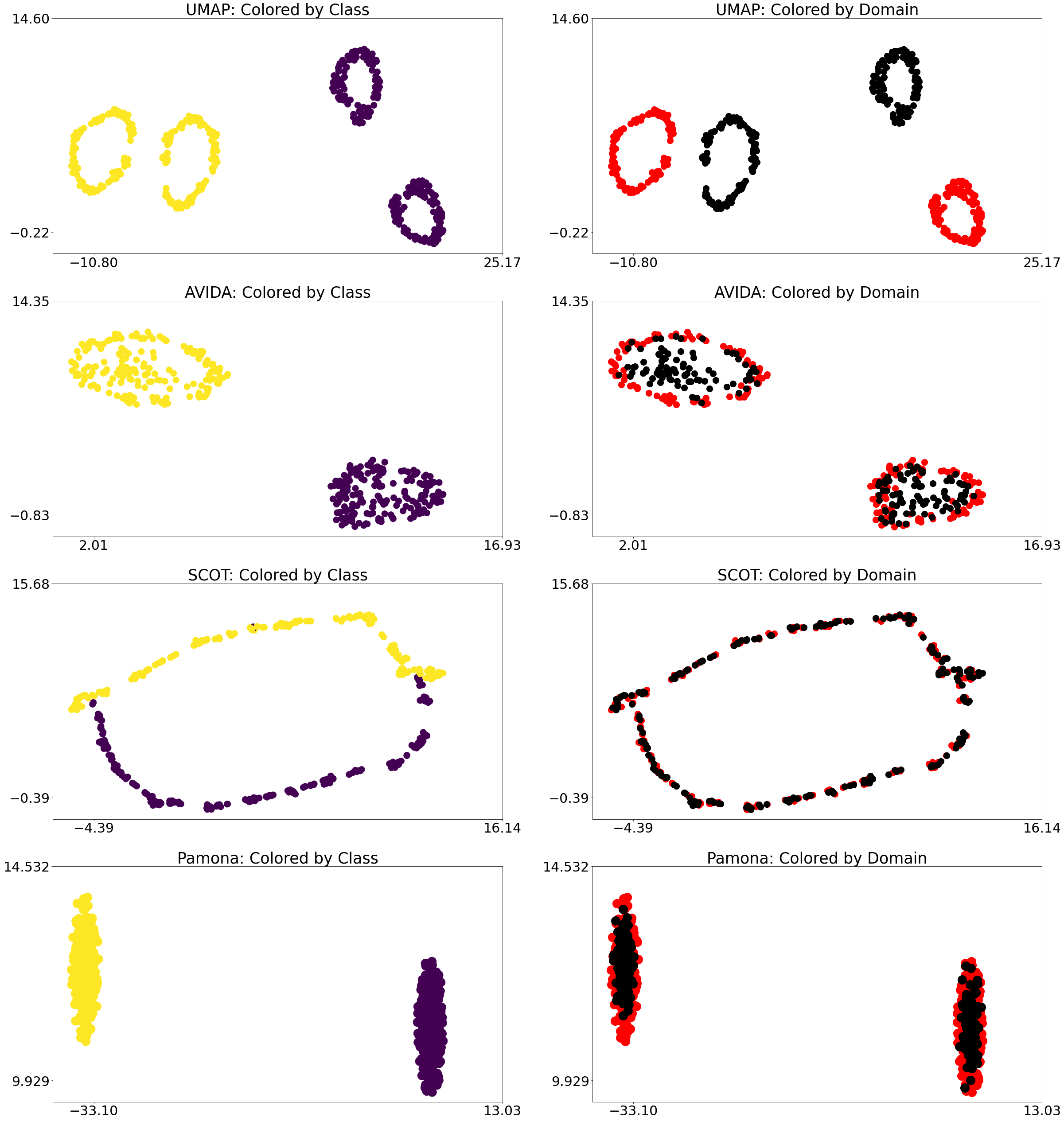}
    \caption{{UMAP, AVIDA, SCOT and Pamona representation of the distant rings dataset. In this experiment, UMAP was applied to SCOT and Pamona's output and UMAP's gradient was incorporated into AVIDA for the dimension reduction module.} }
    \label{fig:umap_rings}
\end{figure*}

 {
In Figure~\ref{fig:umap_rings} we apply $\text{AVIDA}(X_1, X_2; \text{UMAP}, \text{GW})$ to the rings data set described in Section~\ref{sec:data} and see that using UMAP does not preserve the local structure as well as using t-SNE, as shown in Figure~\ref{fig:two_dist_circles}, for all three of the data integration methods. It is not surprising different dimensionality reduction techniques for the same dataaset will produce different representations and we encourage any users of AVIDA to incorporate the dimensionality reduction technique that works best on the dataset they are working with.}
\end{document}